\documentclass[11pt]{article}

\usepackage[final]{acl}
\usepackage{times}
\usepackage{latexsym}

\usepackage[T1]{fontenc}
\usepackage[utf8]{inputenc}

\usepackage{microtype}

\usepackage{inconsolata}

\usepackage{multirow}
\usepackage{graphicx}
\usepackage{amsmath}
\usepackage{booktabs}
\usepackage{algpseudocode}
\usepackage{setspace}
\usepackage{tcolorbox}
\usepackage{amssymb}
\usepackage{xspace}
\usepackage[table]{xcolor}
\definecolor{HeaderGray}{gray}{0.92}
\definecolor{ColGray}{gray}{0.95}
\definecolor{RowGray}{gray}{0.97}
\usepackage{listings}
\tcbuselibrary{breakable}
\newtcolorbox{promptbox}[1][]{
  colback=gray!5,
  colframe=gray!50,
  boxrule=0.4pt,
  left=4pt, right=4pt, top=4pt, bottom=4pt,
  fontupper=\small,
  breakable,
  title=#1
}

\newcommand{\edit}[1]{#1}

\usepackage{enumitem}
\setenumerate[1]{itemsep=0pt,partopsep=0pt,parsep=0pt,topsep=0pt,leftmargin=0pt}
\setitemize[1]{itemsep=0pt,partopsep=0pt,parsep=0pt,topsep=0pt,leftmargin=12pt}
\setdescription{itemsep=0pt,partopsep=0pt,parsep=0pt,topsep=0pt,leftmargin=0pt}

\newtcolorbox{problemBox}{
  colback=gray!10,
  colframe=gray!80,
  boxrule=0.5pt,
  arc=3pt,
boxsep=2pt,
left=0pt, right=0pt, top=0pt, bottom=0pt,
  width=\columnwidth,        % 单栏宽（ACL 双栏里刚好一栏）
  before skip=10pt,
  after skip=10pt,
  fontupper=\normalsize\setstretch{1.05}
}

\newcommand{\systemName}{\textsc{See2Refine}\xspace}

\title{\systemName: Vision-Language Feedback Improves LLM-Based \\ eHMI Action Designers}

\author{
 \textbf{Ding Xia\textsuperscript{1}},
 \textbf{Xinyue Gui\textsuperscript{1}},
 \textbf{Mark Colley\textsuperscript{1,2}},
 \textbf{Zhongyi Zhou\textsuperscript{3}},
 \textbf{Fan Gao\textsuperscript{1}} \\
 \textbf{Dongyuan Li\textsuperscript{1}\thanks{Corresponding author.}},
 \textbf{Renhe Jiang\textsuperscript{1}},
 \textbf{Takeo Igarashi\textsuperscript{1}}
\\
 \textsuperscript{1}The University of Tokyo, \quad
 \textsuperscript{2}University College London, \quad
 \textsuperscript{3}Google
\\
\texttt{dingxia1995@gmail.com},\quad 
\texttt{gui-xinyue@g.ecc.u-tokyo.ac.jp}, \\
\texttt{m.colley@ucl.ac.uk},\quad
\texttt{zhongyizhou@google.com},\quad
\texttt{fangao0802@gmail.com}\\
\texttt{\{lidy, jiangrh\}@csis.u-tokyo.ac.jp},\quad
\texttt{takeo@acm.org}
}

\begin{document}
\maketitle

\begin{abstract}

Automated vehicles lack natural communication channels with other road users, making external Human-Machine Interfaces (eHMIs) essential to convey intent and maintain trust in shared environments. However, most eHMI studies rely on developer-crafted message-action pairs, which are difficult to adapt to diverse and dynamic traffic contexts. A promising alternative is to use Large Language Models (LLMs) as action designers that generate context-conditioned eHMI actions, yet such designers lack perceptual verification and typically depend on fixed prompts or costly human-annotated feedback for improvement.
We present \systemName, a human-free, closed-loop framework that uses vision-language models (VLMs) for perceptual evaluation as automated visual feedback to improve an LLM-based eHMI action designer. Given a driving context and a candidate eHMI action, the VLM evaluates the perceived appropriateness of the action, and this feedback is used to iteratively revise the designer's output, enabling systematic refinement without human supervision.
We evaluate our framework across three eHMI modalities (lightbar, eyes, and arm) and multiple LLM model sizes. Across settings, our framework consistently outperforms prompt-only LLM designers and manually specified baselines in both VLM-based metrics and human-subject evaluations. 
% Results further indicate that the improvements generalize across modalities and that VLM evaluations are well aligned with human preferences, supporting the robustness and effectiveness of \systemName for scalable action design.
\edit{The results further indicate that the improvements are generalized across modalities and that VLM evaluations are reasonably aligned with human preferences in our controlled settings, supporting the robustness and effectiveness of \systemName for scalable action design.}
\footnote{The source code, prompts, and Blender scenarios are available at \url{https://github.com/ApisXia/see2refine}}

\end{abstract}
\section{Introduction}

\begin{figure}[t]
    \centering
    \includegraphics[width=\linewidth]{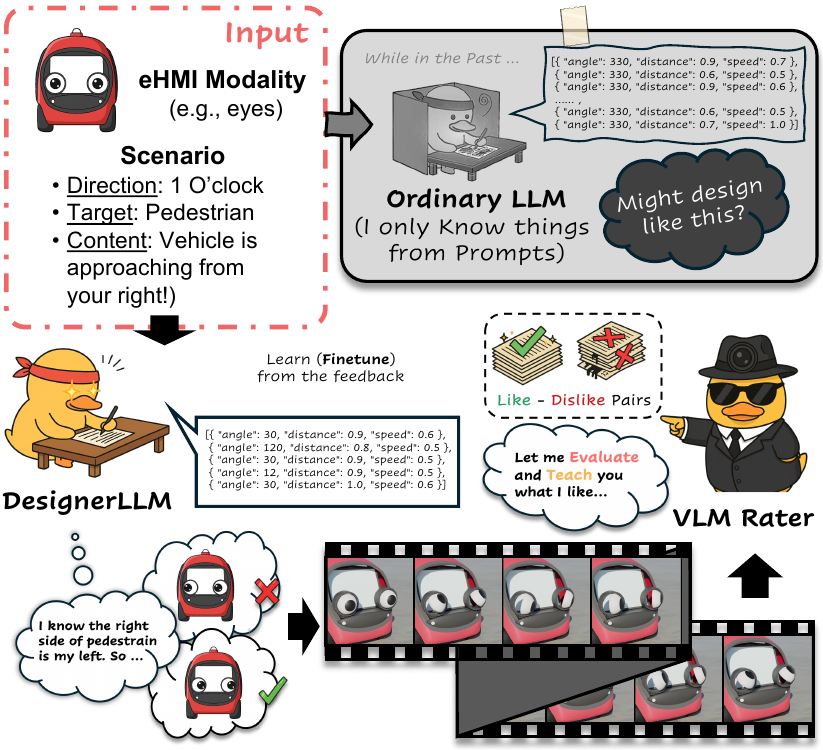}
    \caption{\systemName uses VLM-based perceptual evaluation as automated visual feedback to design, evaluate, and iteratively refine LLM-based eHMI action designers without human supervision. In contrast, standard LLM-based designers rely on static prompts and lack perceptual grounding for improvement.}
    \label{fig:teaser}
\end{figure}

Automated vehicles (AVs) are expected to reach Level 4 autonomy within a decade~\cite{agrawal2023building,chen2023milestones}. Without human drivers, AVs usually lack effective communication channels with other road users, such as pedestrians, cyclists, and human drivers~\cite{fagnant2015preparing}. This gap causes confusion and distrust, leading to series traffic issues~\cite{colley2025improving}. \underline{E}xternal \underline{H}uman-\underline{M}achine \underline{I}nterfaces (eHMIs) are emerging as promising solutions to address this communication challenge~\cite{dey2020taming}. eHMIs can comprise devices such as mechanical eyes~\cite{chang2022can}, mechanical arms~\cite{gui2024shrinkable}, and lightbars~\cite{dey2020color} mounted on the exterior of AVs. These devices serve as communication interfaces that convey messages like ``I am stopping''. For instance, an eHMI can effectively express directional intent through eye gazing, finger pointing, or illuminating particular sections of a lightbar~\cite{gui2024scenarios}. 
% add gui2022going, later
% Current eHMI research explores a range of modalities, from traditional ones, such as lightbar, to anthropomorphic features like eyes and arms~\cite{chang2022can,gui2024shrinkable}, as well as parameter exploration for individual modalities~\cite{lau2022one,colley2025improving}. 

% However, many eHMI actions are still pre-designed by developers, which can limit their adaptability in dynamic and unpredictable environments~\cite{dey2020taming}.

% With eHMIs, researchers design eHMI interactions via \textit{predefined actions}, which limits their adaptability in dynamic and unpredictable environments~\cite{dey2020taming,de2022external}.
% % For example, \xxx.
% % Recently, Large Language Models (LLMs) have demonstrated remarkable versatility and transferability across multiple domains~\cite{radford2019language}.
% % A recent study highlights that Large Language Models (LLMs) can automate eHMI action design in dynamic environments~\cite{xia2025automating}, leveraging their remarkable versatility and transferability across multiple domains~\cite{radford2019language}. 
% LLMs could automate eHMI action design in dynamic environments~\cite{xia2025automating}, as their semantic understanding and generative capabilities enable them to map traffic scenarios directly to appropriate action sequences, eliminating the need for manually predefined rules~\cite{radford2019language}.

With eHMIs, researchers typically design eHMI interactions via predefined actions, limiting adaptability in dynamic and unpredictable environments~\cite{dey2020taming,de2022external}. To address this rigidity, LLMs offer a data-driven alternative by generating context-dependent eHMI actions from traffic scenarios, reducing the reliance on manually defined rules~\cite{radford2019language}. Moreover, previous work shows that, when equipped with carefully crafted system prompts, LLM-based action designers can perform at a level comparable to that of human designers~\cite{xia2025automating}. 
However, text-based prompts can only capture a portion of the details involved in eHMI installation and are unable to convey rich visual information, such as sizes, speeds, and observation angles, which are essential for designing actions in dynamic scenarios~\cite{tellex2020robots, cao2024worst, majumder2024exploring}.

% Reinforcement learning (RL), on the other hand, allows eHMIs to learn what actions to perform without researchers demonstrating them explicit examples.
% Despite its success in related literature in building a generalizable action model~\cite{xxx, xxx}, standard RL requires human feedback data at scale~\cite{kaufmann2023survey,jin2023data,wang2025worldpm}, which does not exist in the eHMI literature, and is expensive to collect. 
% These challenges motivates to explore the following research question:
% \begin{quote}
%     \textbf{RQ1}. How can we build a generalizable LLM to perform eHMI action with an affordable human feedback budget?
% \end{quote}

Using human feedback to guide the extraction of knowledge related to specific types of eHMI serves as a viable solution. Reinforcement Learning from Human Feedback (RLHF)~\cite{bai2022training} has proven effective in developing LLMs at the expert-level in specific tasks. However, applying RLHF to LLM-based eHMI action designers requires extensive human annotations, similar to other RLHF studies~\cite{kaufmann2023survey, jin2023data, wang2025worldpm}. Reinforcement Learning from AI Feedback (RLAIF) offers an alternative approach by leveraging visual perception capabilities of Vision-Language Models (VLMs) to provide judgments that are close to human-level performance~\cite{lee2024prometheus,lu2024wildvision}.
To develop a framework capable of designing, evaluating, and self-improving in a cost-effective, rapid, and automated manner, we pose the following research question:

\begin{center}
\begin{problemBox}
\begin{quote}
    % \textbf{RQ1}: How can we design a system that improves the \textbf{DesignerLLM} for eHMIs through feedback without relying on human annotations?
    \textit{How to use VLMs' perception feedback to improve an LLM-based eHMI action designer without human annotations?}
\end{quote}
\end{problemBox}
\end{center}

In this work, we propose \systemName that enables LLM designers to autonomously design, evaluate, and iteratively refine eHMI actions, utilizing visual perception from VLMs as automated feedback in a cost-effective manner.
\edit{We position this work as an early-stage, controlled-scenario study of simulation-based eHMI refinement.}
In our experiments, we trained separate DesignerLLM models for each of the three eHMI modalities: lightbar, eyes, and arm. 
After three rounds of iterative learning, the DesignerLLM models exhibited a clear alignment in preferences with VLM raters. Additionally, our sampling method to expand the action database effectively reduces annotation costs and time without compromising performance. Subsequently, we recruited 18 participants to evaluate the eHMI actions generated by five LLMs: two 7B models (one base model, one DesignerLLM) and three state-of-the-art models. Extensive results demonstrate that leveraging the visual perception capabilities of VLMs can also enhance human evaluation scores. 
The main contributions of this study can be summarized as follows:
\edit{
\begin{itemize}
    \item We propose \systemName, a closed-loop framework that integrates VLM-based perceptual evaluation as automated visual feedback to improve LLM-based eHMI action designers in controlled simulated settings.
    \item We discuss an efficient database expansion strategy that combines importance-based scenario sampling and diverse action generation to improve refinement efficiency without sacrificing preference alignment.
    \item We show through a human study that VLM-guided refinement can transfer to measurable improvements in human-perceived eHMI quality across the evaluated scenarios.
\end{itemize}
}

\section{Related Work}
% \subsection{eHMI Action Planning}
% \subsubsection{Rule-Based Action Planning}
% Existing research on eHMI action planning typically relies on predetermined designs, where human designers define behavioral rules tailored to specific eHMI modalities. For text- and icon-based systems, designers base message content on conventional traffic symbols or standard textual cues~\cite{eisele2022effects,eisma2021external}. Color- and light-based eHMIs are designed through intuitive color associations and empirical testing of flashing patterns~\cite{bazilinskyy2019survey, dey2020color}. Anthropomorphic designs, such as those featuring eyes or arm gestures, draw on nonverbal human communication principles~\cite{mahadevan2018communicating,ochiai2011homunculus}. 
% % Most recently, \cite{colley2025improving} proposes using Human-In-The-Loop Multi-Objective Bayesian Optimization to create appropriate eHMIs. 
% Despite these innovations, most approaches remain labor-intensive, difficult to generalize, and often lack scalability for complex real-world scenarios~\cite{gui2023field,de2022external}.
% Recent advances in LLMs present a promising shift in eHMI design. Xia et al.~\cite{xia2025automating} demonstrate that, when provided with detailed scenario descriptions and target messages, LLMs can generate eHMI actions approaching human-level quality. However, their zero-shot performance presents challenges in prompt formulation and remains constrained by a limited understanding of nuanced visual contexts and the specific semantics inherent to eHMI environments.

\paragraph{eHMI Action Planning.} Existing eHMI action-planning mostly follows predefined, designer-authored rules per modality: text/icons use traffic-rules messages~\cite{eisele2022effects,eisma2021external}; color/light use flashing patterns based on intuitive color associations~\cite{bazilinskyy2019survey,dey2020color}; and anthropomorphic cues (e.g., eyes, arm gestures) draw on nonverbal communication principles~\cite{mahadevan2018communicating,ochiai2011homunculus}. Although effective, these manual designs are labor-intensive and hard to scale to complex real-world scenarios~\cite{gui2023field,de2022external}. Recent work uses LLMs for eHMI action planning via prompt engineering~\cite{xia2025automating}, but this remains prompt-dependent and largely task-level, lacking mechanism-level understanding, requiring more advanced methods for higher-quality action planning.

\begin{figure*}[ht!]
    \centering
    \includegraphics[width=\linewidth]{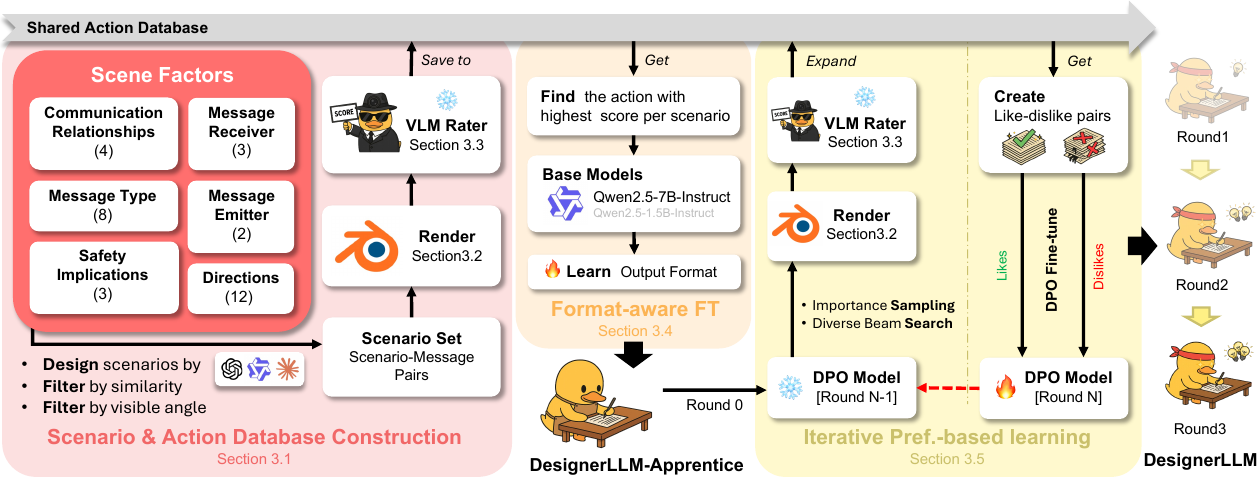}
    \caption{Our \systemName framework includes: scenario and action database construction (Section~\ref{method:data}), format-aware fine-tuning (Section~\ref{method:sft}), and iterative preference-based learning (Section~\ref{method:dpo}). A shared action database supports all three components, storing generated actions and expanding to enhance DesignerLLM’s performance.}
    \label{fig:overall_pipeline}
\end{figure*}

\paragraph{VLM-based perceptual evaluation as feedback.} Reinforcement Learning from AI Feedback (RLAIF) replaces the collection of human preferences in RLHF by using stronger teacher models to generate reward signals~\cite{bai2022constitutional,lee2023rlaif}. In multimodal settings, VLMs can be fine-tuned as rubric-following judges for images and videos, achieving human-level agreement and performance competitive with GPT-4V~\cite{lu2024wildvision,chen2024mllm}. Systems such as Prometheus-Vision, WildVision-Arena, WildVision-Bench, and MLLM-as-a-Judge show that VLM judges can follow instructions and output pointwise or pairwise scores that align with human preference orderings~\cite{lee2024prometheus,lu2024wildvision,chen2024mllm}, while VHELM and LLaVA-Critic further standardize multimodal evaluation and reduce reliance on closed models~\cite{lee2024vhelm,xiong2025llava}. Multimodal variants of RLAIF and RLHF-V use AI-generated preferences to train reward models and optimize large VLM policies with minimal human labeling~\cite{ahn2024tuning,yu2025rlaif,sun2023aligning}. For eHMI, VLM judges correlate well with human ratings on LLM-generated action clips, enabling closed-loop optimization where an LLM proposes actions, a VLM scores them, and updates (e.g., policy gradient or DPO-style) refine the space without human studies~\cite{wang2024rl}.
\section{Method}

As shown in Figure~\ref{fig:overall_pipeline}, our \systemName framework contains five components, including (1) Scenario-Message Pair Generation, (2) Model Asset \& Action Rendering, (3) Multi-Metric Evaluation System, (4) Format-aware Fine-tuning, and (5) Iterative Preference-based Learning. In the following, we describe each component in detail.

\begin{figure*}[t]
    \centering
    \includegraphics[width=\linewidth]{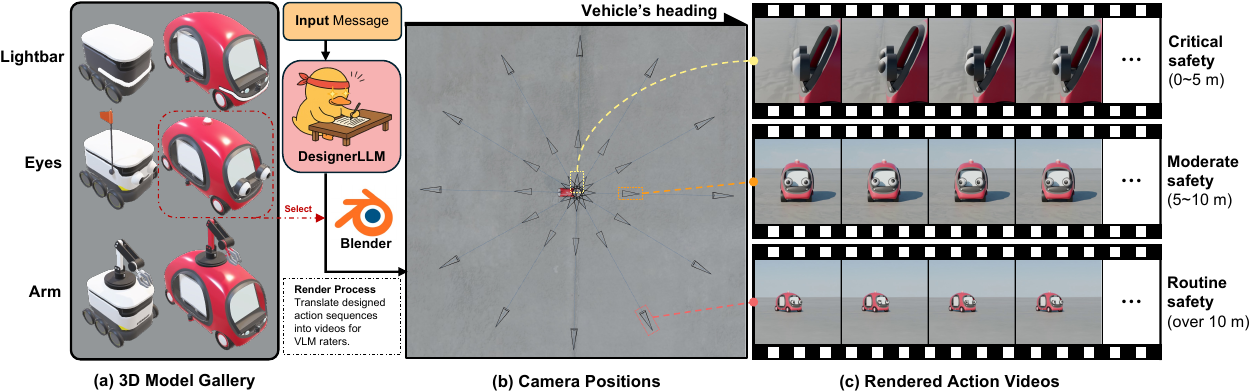}
    \caption{Overview of the six eHMI 3D models combining three modalities (lightbar, eyes, arm) and two emitter types (self-driving car, delivery robot). Rendered videos are generated in Blender from the message receivers’ perspective under a defined camera direction and distance.}
    \label{fig:model_and_blender}
\end{figure*}

\subsection{Scenario-Message Pair Generation}
\label{method:data}
% The section involves two aspects: generating traffic scenario-message pairs and creating an initial action database, a
Inspired by existing eHMI scenario studies~\cite{dey2020taming,colley2020design}, we first generate traffic scenarios by combining six different factors. The factors we consider are listed below. By combining these factors, we generate a total of 6,912 unique condition combinations.

\begin{itemize}%[leftmargin=*]
    \item \textbf{Communication Relationships (4):} 1st-person and 3rd-person perspective $\times$ one-to-one and one-to-many.
    \item \textbf{Emitter (2):} Self-driving car; delivery robot.
    \item \textbf{Receiver (4):} Vehicle driver; pedestrian; cyclist; motorcyclist.
    \item \textbf{Message Type (8):} Instruction, advisory, question, answer, current, historical, predictive, and warn.
    \item \textbf{Direction (12):} 1 to 12 o'clock.
    \item \textbf{Safety Implications (3):} Critical (0--5 m), moderate (5--10 m), and routine ($>$10 m).
\end{itemize}

Second, conditioned on these scenarios, we generate 20,736 intended messages containing scenario information using larger LLMs, including OpenAI GPT-4.1~\cite{openai2025gpt41}, Claude 3.7 Sonnet~\cite{anthropic2025claude37}, and Qwen3-235B-A22B~\cite{qwen2025qwen3235ba22b}. To eliminate redundancy, we employ a sentence transformer~\cite{reimers2019sentence} to compare messages. We sample 70\% of these messages using the farthest point sampling algorithm~\cite{eldar1997farthest} to ensure sufficient variance. 
In addition, we conduct a small user-rating study on 200 randomly selected scenario–message pairs. Two eHMI researchers independently rate the validity of each pair on a 7-point Likert scale (1=Strongly Disagree, 7=Strongly Agree). The mean rating is 5.3 (SD=1.9), between ``Somewhat Agree'' (5) and ``Agree'' (6), suggesting that the generated scenarios are generally of high quality.

Third, we build an initial action database for the later co-learning process. Using the same LLMs as in the second step, we generate two structured action sequences per model for each modality (six candidates per scenario). We then render these actions (Section~\ref{method:modality}) and score them with VLM raters (Section~\ref{method:metrics}), saving the scored actions to the shared database.

\subsection{Model Asset and Action Rendering}
\label{method:modality}

\subsubsection{Model Development and Definition}
In this study, we focus on three eHMI modalities: lightbar, eyes, and arm. The first two modalities represent classic eHMI designs~\cite{dey2020taming,deb2018investigating,rouchitsas2019external,benderius2017best,colley2020towards} with simpler control mechanisms, while the arm modality requires managing more parameters, which presents additional challenges for \textbf{DesignerLLM}. In Blender~\citep{blenderorg}, we implement each modality in two message emitter types (self-driving cars and delivery robots), resulting in six 3D models, as shown in \autoref{fig:model_and_blender} (a). Modality descriptions and permitted actions are as follows:

\begin{itemize}
    \item \textbf{Eyes} are mounted on the front of the vehicle. The pupil is parameterized in polar coordinates: angle in $[0^\circ,360^\circ]$ (0$^\circ$=up, counterclockwise) and radius in $[0,1]$ (0=center, 1=edge)~\cite{chang2022can,gui2022going}.
    \item \textbf{Light Bar} is mounted on the vehicle front, slightly extending to both wings. It has 16 binary lighting regions (0=off, 1=on); e.g., ``0000000011111111'' denotes left-half off and right-half on (vehicle perspective). The brightness is fixed and the color is cyan~\cite{dey2020color}.
    \item \textbf{Arm} is mounted on the vehicle top and consists of five single-axis joints (shoulder, upper arm, forearm, hand, fingers) operating within limited ranges~\cite{gui2024shrinkable}.
\end{itemize}

We also specify transition speeds between action states. The eye and arm transitions range from 0.5 to 2.0s (0.1s steps), while the light bar transitions range from 0.1 to 1.0s (0.1s per step). This enables smooth sequences while preserving temporal detail in the video FPS.

\subsubsection{Rendering Pipeline}
\label{method:render}
For rendering, we typically consider three conditions: camera direction, camera distance (relative to the object), and background scenes, as illustrated in \autoref{fig:model_and_blender} (b). Regarding the direction of the camera, aligned with Section~\ref{method:data}, we create 12 cameras arranged clockwise. Regarding the camera distance, which reflects the safety implications (Section~\ref{method:data}), we position the cameras at three different distances. 
These positions can be accessed via Python to generate videos that simulate the perspectives of other road users. 
We use a plain background, unlike the previous work~\cite{xia2025automating}, to avoid contextual cues and reduce rendering time. Removing scene backgrounds decreases the rendering time to approximately $1/10$, while lower GPU memory usage also enables more parallel threads, further reducing wall-clock time by around 4× for batch rendering.

We render action videos using Blender version 4.5~\cite{blenderorg} on a GPU-equipped device featuring four NVIDIA GTX Titan cards with 24 GB of RAM each, allowing up to 16 concurrent threads. The average rendering time for 1,000 clips is approximately one hour, with each thread consuming around 1.1 GB of GPU RAM. 
Each frame is rendered at a resolution of 512×512 pixels, centered on the AV equipped with eHMI, as shown in \autoref{fig:model_and_blender} (c). These clips are rendered at four FPS, with a maximum action change rate for each eHMI set to ensure that the frame rate captures all relevant details, following the Nyquist-Shannon sampling theorem~\cite{nyquist1928certain}.

\subsection{Multi-Metric Evaluation System}
\label{method:metrics}
Recent studies show that VLMs demonstrate capabilities and preferences similar to those of humans~\cite{lee2024prometheus,lu2024wildvision,chen2024mllm}. In our task, our aim is to borrow knowledge from existing Human-Computer Interaction research~\cite{colley2020towards,colley2020design,colley2025improving} and use the features of VLMs to develop a multi-metric evaluation system.

To simulate the human perception process, we design a two-phase assessment: Phase 1 (without revealing the intended message) and Phase 2 (with the intended message disclosed). Each phase assesses different aspects using standardized 9-point Likert rating scales. 

Phase 1 assesses how effectively the eHMI communicates without prior knowledge of the message, simulating real-world reception scenarios.
\begin{itemize}
    \item \textbf{Intention Recognition (text + certainty, 9-pt):} Infer the message from the animation only; output (i) an interpreted message sentence and (ii) a 9-point certainty score (1=unclear, 5=neutral, 9=unambiguous).
    \item \textbf{Targeting (9-pt):} Rate the confidence that the message is directed to the receiver (1=not for the receiver, 5=uncertain, 9= clear for the receiver).
    \item \textbf{Trust (9-pt):} Rate trust in AV/eHMI (1=distrust/avoid, 5=neutral, 9=full trust).
    \item \textbf{Similarity (9-pt, post-hoc):} Compare the VLM-interpreted message with the intended message (from the emitter scenario) using a small LLM judge (1=contradictory, 5=partially similar, 9=equivalent).
\end{itemize}

Phase 2 reveals the intended message to the VLM evaluator and assesses how well the observed eHMI behavior aligns with this communication goal. This phase evaluates the quality of the design from an informed perspective.
\begin{itemize}
    \item \textbf{User Acceptance (9-pt):} Willingness to accept the eHMI in daily life given the scenario, intended message, and actions (1=reject, 5=uncertain, 9=fully accept).
    \item \textbf{Consistency (9-pt):} Alignment between the perceived meaning of the actions shown and the revealed intended message (1=contradictory, 5=mixed, 9=fully aligned).
\end{itemize}

\subsubsection{Kernel Score: Composite Quality Metric}

We later define a kernel scoring function that aggregates the evaluation metrics as a unified quality indicator. Given the six scores obtained from the VLM evaluation: target score $t$, user acceptance $u$, consistency $c$, certainty $\kappa$, similarity $s$, and trust $\tau$. We compute the kernel score $K$ as:

\begin{equation}
K = (\kappa \times s) + t + \tau + u + c.
\end{equation}

\edit{We use this kernel score as a simple default aggregation rather than as a claim of task-independent optimal weighting.}

\subsection{Format-aware Fine-tuning}
\label{method:sft}
This phase trains \textbf{DesignerLLM-Apprentice} to reliably output eHMI-specific action formats, as shown in Figure~\ref{fig:overall_pipeline}. We fine-tune two base model sizes, Qwen2.5-7B-Instruct and Qwen2.5-1.5B-Instruct~\cite{qwen2024qwen25} as ``apprentices'' for later optimization.
Training data is sampled from the shared action database by selecting the candidate action with the best kernel score per scenario. 

\subsection{Iterative Preference-based learning}
\label{method:dpo}
We utilize an iterative Direct Preference Optimization (DPO) approach~\cite{du2024blenderllmtraininglargelanguage,rafailov2023direct} to progressively enhance the ability of DesignerLLM-base models to generate high-quality eHMI action designs. Our pipeline involves multiple cycles of sampling, rendering, evaluation, and fine-tuning.

\subsubsection{Importance-based Scenario Sampling}
Expanding the action database is common in DPO-style iterative training~\cite{du2024blenderllmtraininglargelanguage}, but regenerating actions for all scenarios is costly, and random sampling can overlook hard cases. We therefore use importance sampling to prioritize scenarios that benefit most from additional generations.

For each scenario $i$, the shared action database stores tuples $(i, m, K_i^m)$, where $m$ indexes the source model/round and $K_i^m$ is the kernel score. From these records, we compute
$K_i^{\text{best}}=\max_m K_i^m$, $K_i^{\text{worst}}=\min_m K_i^m$,
$\Delta K_i=K_i^{\text{best}}-K_i^{\text{worst}}$, and the candidate count $N_i$.
We define the importance score as follows:
\begin{equation}
I_i = \frac{(\Delta K_{\max} - \Delta K_i)\, K_i^{\text{worst}}}{(K_i^{\text{best}})^3}\times 0.5^{n_i},
\end{equation}
where $\Delta K_{\max}=\max_j \Delta K_j$ and $n_i=N_i/6-1$ (assuming 6 generations of baseline).
It prioritizes:
\begin{itemize}
    \item Low best scores $K_i^{\text{best}}$ (room for improvement);
    \item Small gaps $\Delta K_i$ (limited diversity);
    \item High worst scores $K_i^{\text{worst}}$ (easier to improve);
    \item Fewer previous sampling attempts. 
\end{itemize}
We normalize scores such that $\max_i I_i=1$, then sample 20\% of scenarios per round by default.

\subsubsection{Diverse Action Generation}
For each sampled scenario, we generate diverse eHMI action sequences with the current fine-tuned model using diverse beam search~\cite{vijayakumar2016diverse} (six beam groups with one beam each). Each output is validated against modality-specific JSON schema (eyes/arm/lightbar) via Pydantic, and invalid formats are discarded.
We then render and evaluate the valid actions using the same pipeline as the initial database, compute their kernel scores, and append them to the shared action database. Repeating this step iteratively expands the candidate set and improves both diversity and quality over training rounds.

\subsubsection{Preference Pair Construction}
After expanding the shared score database, we form DPO preference pairs $(y^+,y^-)$ per scenario, where $y^+$ is a higher-quality action than $y^-$. This setup is analogous to the ordering of human preferences, as discussed in previous work~\cite{xia2025automating}. Pair extraction is two-stage. (1) \textbf{Max--min:} for each scenario $i$, select the highest and lowest-scoring actions,
\begin{equation}
(y_i^+,y_i^-)=(\arg\max_m K_i^m,\ \arg\min_m K_i^m),
\end{equation}
only if the score gap $\Delta K_i \ge \delta_{\min}$ (default $\delta_{\min}=4.0$), to ensure a strong preference signal. (2) \textbf{High-gap extras:} collect additional pairs within the scenario with $K_i^{m_1}-K_i^{m_2}\ge\delta_{\min}$, sort by gap, and keep the top $p\%$ (default $30\%$) to enhance training.

\section{Experiment}
% We evaluate the framework in two parts: first, by quantitatively analyzing preference alignment after the co-learning process and the effectiveness of various modules

\subsection{Experimental Settings}
\label{exp:setup}

\paragraph{Devices and Software.}
LLM inference and video rendering run locally on a 4$\times$RTX TITAN (24GB GPU RAM each) machine. Fine-tuning (Format-aware Fine-tuning; Iterative Preference-based learning) is performed on cloud GPUs with up to 2$\times$A100 (40GB GPU RAM each). We use Transformers for inference~\cite{wolf-etal-2020-transformers} and LLaMA-Factory for fine-tuning~\cite{zheng2024llamafactory}. VLM inference is performed remotely using GPT-5-mini~\cite{openai2025gpt41}.

\paragraph{Training Setup.}
All fine-tuning processes use ShareGPT-formatted data~\cite{vicuna2023} with an 80/20 train/test split. We fine-tune with LoRA~\cite{hu2022lora} in bfloat16: 3 epochs for Format-aware Fine-tuning and 1 epoch for DPO (to reduce overfitting, similar to previous works~\cite{bai2022constitutional,touvron2023llama,liu2024deepseek}). DPO uses weight=1, sigmoid loss, and a preference coefficient of 0.01.

\paragraph{Computation Cost.}
VLM rating achieves approximately 90 evaluations per second with 8 threads. Single-thread inference requires about 7 seconds per scenario for the 7B model and 6 seconds per scenario for the 1.5B model. The rendering takes roughly 45 seconds per clip, operating in batch parallelization. Fine-tuning duration is approximately 3 hours for the 7B model and 1 hour 10 minutes for the 1.5B model in Format-aware Fine-tuning, and around 5 hours (7B) or 3 hours 30 minutes (1.5B) per DPO round. Excluding dataset construction and assuming three DPO rounds, the total co-learning time amounts to at least 60 hours for the 7B model and 38 hours for the 1.5B model.
\edit{Importantly, this rendering-and-evaluation cost is incurred only during offline refinement. At deployment time, \systemName uses the trained DesignerLLM to generate actions directly, without requiring video rendering or VLM-based evaluation.}

\subsection{Results}

\begin{table*}[!t]
\centering
\footnotesize
\resizebox{\textwidth}{!}{%
\begin{tabular}{llccccccc}
\toprule
\multirow{2}{*}{\textbf{Modality}} & \multirow{2}{*}{\textbf{Source / Configuration}} & \multicolumn{6}{c}{\textbf{VLM rater Metrics}} & \multirow{2}{*}{\textbf{Div.~$\uparrow$}}\\
\cmidrule(lr){3-8} & & \textbf{UA~$\uparrow$} & \textbf{Cons.~$\uparrow$} & \textbf{Targ.~$\uparrow$} & \textbf{Trust~$\uparrow$} & \textbf{Sim.~$\uparrow$} & \textbf{K.S.~$\uparrow$} & \\
\midrule
\multirow{7}{*}{Lightbar} 
& DesignerLLM-7B$^{\dagger}$ (Full) & \cellcolor{HeaderGray}\textbf{5.889} & \cellcolor{HeaderGray}\textbf{6.061} & \cellcolor{HeaderGray}\textbf{6.577} & \cellcolor{HeaderGray}\textbf{6.874} & \cellcolor{HeaderGray}0.373 & \cellcolor{HeaderGray}\textbf{28.017} & \cellcolor{HeaderGray}63.580 \\
& DesignerLLM-1.5B$^{\dagger}$ & 5.725 & 5.898 & 6.516 & 6.865 & 0.351 & 27.473  & \textbf{63.888} \\
& DesignerLLM-Apprentice-7B$^{\dagger}$ & 4.847 & 4.763 & 6.273 & 6.796 & \textbf{0.377} & 25.382 & 61.974 \\
& Qwen2.5-7B-Instruct$^{*}$ & 4.382 & 4.125 & 5.909 & 6.705 & 0.362 & 23.759 & 53.441 \\
& Initial Action Database$^{\ddagger}$ & 4.444 & 4.203 & 6.197 & 6.784 & 0.365 & 24.244 & 58.378 \\
\cmidrule(lr){2-9}
& \textit{w/o Div.Beam} & 5.011 & 5.234 & 6.377 & 6.806 & 0.370 & 24.534 & 56.332 \\
& \textit{w/o Imp.Samp.} & 5.097 & 5.386 & 6.439 & 6.839 & 0.372 & 24.884 & 54.034 \\

\midrule
\multirow{7}{*}{Eyes} 
& DesignerLLM-7B$^{\dagger}$ (Full) & \cellcolor{HeaderGray}\textbf{6.010} & \cellcolor{HeaderGray}\textbf{6.083} & \cellcolor{HeaderGray}\textbf{6.929} & \cellcolor{HeaderGray}\textbf{6.925} & \cellcolor{HeaderGray}0.432 & \cellcolor{HeaderGray}\textbf{28.839} & \cellcolor{HeaderGray}8.492 \\
& DesignerLLM-1.5B$^{\dagger}$ & 5.828 & 5.919 & 6.773 & 6.901 & 0.421 & 28.246 & 8.229 \\
& DesignerLLM-Apprentice-7B$^{\dagger}$ & 5.461 & 5.533 & 6.598 & 6.813 & 0.400 & 27.158 & 7.242 \\
& Qwen2.5-7B-Instruct$^{*}$ & 5.171 & 5.042 & 6.629 & 6.869 & \textbf{0.435} & 26.642 & \textbf{10.726} \\
& Initial Action Database$^{\ddagger}$ & 5.110 & 5.103 & 6.455 & 6.843 & 0.413 & 26.269 & 7.770 \\
\cmidrule(lr){2-9}
& \textit{w/o Div.Beam} & 5.633 & 5.649 & 6.706 & 6.908 & 0.428 & 27.747 & 7.812 \\
& \textit{w/o Imp.Samp.} & 5.706 & 5.732 & 6.761 & 6.900 & 0.424 & 27.936 & 7.976 \\

\midrule
\multirow{7}{*}{Arm} 
& DesignerLLM-7B$^{\dagger}$ (Full) & \cellcolor{HeaderGray}\textbf{6.569} & \cellcolor{HeaderGray}\textbf{6.974} & \cellcolor{HeaderGray}\textbf{7.354} & \cellcolor{HeaderGray}6.759 & \cellcolor{HeaderGray}\textbf{0.428} & \cellcolor{HeaderGray}\textbf{30.531} & \cellcolor{HeaderGray}14.721 \\
& DesignerLLM-1.5B$^{\dagger}$ & 6.495 & 6.820 & 7.313 & \textbf{6.768} & 0.423 & 30.218 & 14.896 \\
& DesignerLLM-Apprentice-7B$^{\dagger}$ & 6.109 & 6.362 & 7.156 & 6.614 & 0.413 & 29.128 & 15.022 \\
& Qwen2.5-7B-Instruct$^{*}$ & 5.202 & 5.156 & 6.807 & 6.686 & 0.412 & 26.640 & 10.891 \\
& Initial Action Database$^{\ddagger}$ & 5.935 & 6.090 & 7.142 & 6.730 & 0.425 & 28.759 & \textbf{16.282} \\
\cmidrule(lr){2-9}
& \textit{w/o Div.Beam} & 6.359 & 6.622 & 7.237 & 6.732 & 0.417 & 29.766 & 16.920 \\
& \textit{w/o Imp.Samp.} & 6.366 & 6.594 & 7.233 & 6.755 & 0.425 & 29.694 & 16.788 \\
\bottomrule
\end{tabular}}
\caption{Comprehensive VLM evaluation and ablation results. $\ddagger$ denotes the structured prompt with explicit format guidance; $\dagger$ is a simplified version of $\ddagger$ without format guidance; $*$ refers to an enhanced prompt adapted from prior work~\cite{xia2025automating}. For ablation, we show the removal of Scenario Sampling (Imp.Samp.) and Diverse Action Generation (Div.Beam) modules.}
\label{tab:merged_results}
\end{table*}

\subsubsection{Alignment with VLM Rater Preferences.}
We evaluate how well DesignerLLM aligns with VLM rater preferences to validate the effectiveness of our framework. We compare DesignerLLM-7B with DesignerLLM-1.5B, DesignerLLM-Apprentice-7B, the Qwen2.5-7B-Instruct base and the initial action database (Section~\ref{method:data}). We report all learning metrics and a diversity measure, adapted from the Action Reference Score (ARS)~\cite{xia2025automating}, which assesses the average similarity between pairs of action sequences. Notably, these five models use three different prompts, as described in the caption of Table~\ref{tab:merged_results}.

\noindent\textbf{Preference Alignment.}
Table~\ref{tab:merged_results} shows that all DesignerLLM variants outperform the initial action database on all metrics, indicating a successful alignment with VLM raters. Gains are larger for Phase~2 metrics (acceptance, consistency) than Phase~1 metrics (targeting, trust, similarity), with trust and similarity improving the least (also discussed in Section~\ref{exp:metric_analysis}).

% \textbf{Model size.}
% DesignerLLM-7B achieves higher average scores than DesignerLLM-1.5B, suggesting better alignment with larger capacity.

\noindent\textbf{Effect of Multi-Round Learning.}
Comparing the post-SFT base models (DesignerLLM-base) with the final models (DesignerLLM) confirms that iterative preference-based learning is necessary to further distill VLM preferences into DesignerLLM.

\begin{table*}[t]
\centering
\footnotesize
\resizebox{\textwidth}{!}{%
\begin{tabular}{lccccccccc}
\toprule
\multirow{2}{*}{\textbf{Modality}} & \multirow{2}{*}{\textbf{Stage}} &
\multicolumn{6}{c}{\textbf{VLM rater Metrics}} &
\multirow{2}{*}{\textbf{Div.~$\uparrow$}} &
\multirow{2}{*}{\textbf{F.Err. (\%)~$\downarrow$}} \\ 
\cmidrule(lr){3-8}
& & \textbf{UA~$\uparrow$} & \textbf{Cons.~$\uparrow$} &
\textbf{Targ.~$\uparrow$} & \textbf{Trust~$\uparrow$} &
\textbf{Sim.~$\uparrow$} & \textbf{K.S.~$\uparrow$} &  & \\ 
\midrule
\multirow{4}{*}{Lightbar}
& Apprentice & 4.847 & 4.763 & 6.273 & 6.796 & \textbf{0.377} & 25.382 & 61.974 & \textbf{<0.01} \\
& Round1 & 5.673 & 5.888 & 6.526 & 6.826 & 0.352 & 27.425 & \textbf{66.628} & 6.06 \\
& Round2 & \textbf{5.920} & \textbf{6.152} & \textbf{6.637} & 6.848 & 0.359 & \textbf{28.110} & 62.983 & 10.06 \\
& Round3 & 5.889 & 6.061 & 6.577 & \textbf{6.874} & 0.373 & 28.017 & 63.580 & 11.31\\
\midrule
\multirow{4}{*}{Eyes}
& Apprentice & 5.461 & 5.533 & 6.598 & 6.813 & 0.400 & 27.158 & 7.252 & 0 \\
& Round1 & 5.707 & 5.742 & 6.733 & 6.733 & 0.423 & 27.917 & 8.229 & 0\\
& Round2 & 5.834 & 5.921 & 6.790 & 6.912 & 0.427 & 28.305 & \textbf{8.606} & 0 \\
& Round3 & \textbf{6.010} & \textbf{6.083} & \textbf{6.929} & \textbf{6.925} & \textbf{0.432} & \textbf{28.839} & 8.492 & 0 \\
\midrule
\multirow{4}{*}{Arm}
& Apprentice & 6.109 & 6.362 & 7.156 & 6.614 & 0.413 & 29.128 & 15.022 & 0 \\
& Round1 & 6.377 & 6.604 & 7.235 & 6.754 & 0.426 & 29.826 & \textbf{16.458} & 0\\
& Round2 & 6.471 & 6.743 & 7.301 & 6.755 & \textbf{0.432} & 30.168 & 15.197 & 0 \\
& Round3 & \textbf{6.569} & \textbf{6.974} & \textbf{7.354} & \textbf{6.759} & 0.428 & \textbf{30.531} & 14.721 & <0.01\\
\bottomrule
\end{tabular}}
\caption{Different Round Performance and Output Format Error Comparison.}
\label{tab:round_compare}
\end{table*}

\subsubsection{Ablation Studies}
We conduct ablations to (1) test whether Scenario Sampling and Diverse Action Generation improve the speed--performance trade-off, and (2) track DesignerLLM performance across different rounds.

\noindent\textbf{Module Effectiveness.}
We remove each module while fixing the sampling ratio at 20\% to keep training tractable (full regeneration would exceed 13 days). As shown in \autoref{tab:merged_results}, omitting either module causes a clear performance drop, especially for the lightbar modality, where the decline is roughly twice that of the other modalities on average.

\noindent\textbf{Across-Round Behavior.}
We further compare DesignerLLM-7B across 3 rounds (Table~\ref{tab:round_compare}), adding \emph{Formatting Errors} (F.Err.) to quantify output validity. Notably, lightbar shows degraded format accuracy after finetuning, likely because its 16-bit status strings (e.g., ``0000000011111111'') encounter issues with continuous sequences of digits in current LLMs~\cite{spathis2023first,huggingface2024numbertokenization}.

\begin{figure*}[t]
    \centering
    \includegraphics[width=\linewidth]{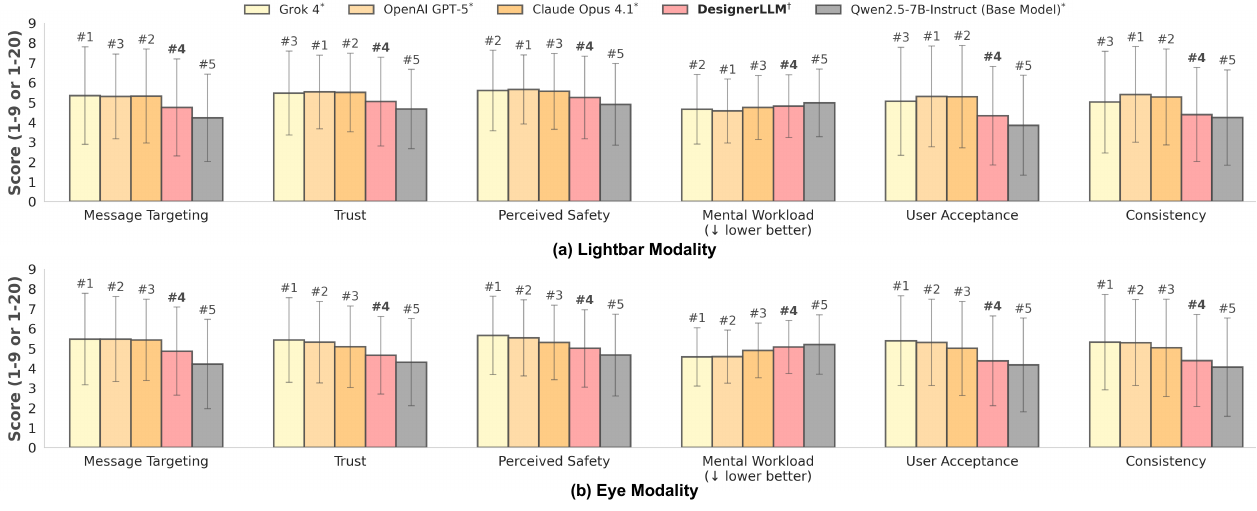}
    \caption{Human ratings of five models on six subjective metrics for (a) eyes and (b) lightbar. Bars show mean scores (1--9) with error bars; mental workload is rescaled from 1--20 to 1--9. Ranks are shown above bars (\#1 best). DesignerLLM (7B) ranks 4th for both modalities, outperforming the base model and approaching commercial models. $\dagger$ denotes a simple eHMI description; $*$ denotes an enhanced prompt adapted from~\cite{xia2025automating}.}
    \label{fig:user_study_result}
\end{figure*}

\subsubsection{Metric-wise Analysis}
\label{exp:metric_analysis}
The metrics choices are inherited from previous eHMI work, but their effectiveness for VLM raters is unclear. Therefore, we analyze the VLM score distributions for three milestones: (1) the initial action database, (2) DesignerLLM-Apprentice-7B, and (3) DesignerLLM-7B.

As illustrated in Figure~\ref{fig:vlm_metric_plot} in the Appendix, metrics behave differently over training. Phase~2 metrics (User Acceptance, Consistency) improve most clearly: their means increase, and distributions tighten, suggesting that they are the most sensitive and reliable signals for VLM-based evaluation.

\edit{In contrast, Phase 1 metrics show weaker trends. Targeting improves slightly, while trust changes little and provides poor preference separation in the initial database, making it less useful for constructing training pairs. We suspect that this reflects two factors. First, trust is a higher-level subjective judgment than acceptance or consistency and may therefore be harder for current VLMs to approximate reliably. Second, our controlled renderings omit environmental cues, such as background context, lighting, and occlusion, which may be particularly important for trust-related judgments. This suggests that \systemName is currently better suited to optimize more directly perceivable communication qualities than nuanced subjective impressions such as trust.}

% \subsection{User Study}
\subsection{Human Preference Alignment}
We conduct a user study to (1) test whether improvements transfer from VLM raters to people, and (2) examine potential gaps between VLM and human preferences. Details are provided in Appendix~\ref{app:user_study}.

\noindent\textbf{Setup.}
We evaluate two common modalities (eyes, lightbar) on eight scenario-message pairs from prior work~\cite{xia2025automating}. We compare five models: DesignerLLM-base, DesignerLLM, and three commercial LLMs (OpenAI GPT-5, Claude Opus 4.1, Grok~4). Each model generated actions that are rendered as 1080p videos at 12 FPS with full scene context (80 videos total).

\noindent\textbf{Participants and measures.}
We recruit 18 participants (21--30 years, $M{=}26$, $SD{=}2.7$), split evenly by modality. For each scenario, participants rate videos in two stages: without the intended message (targeting, trust, perceived safety, mental workload) and with the intended message revealed (acceptance, consistency).

\noindent\textbf{Results.}
In both modalities, DesignerLLM consistently outperform the base model and rank 4th overall (Figure~\ref{fig:user_study_result}). Relative to the base model, DesignerLLM improves the average scores by 7.7\% (eyes) and 7.9\% (lightbar), with the largest gains in targeting (15.4\% eyes; 12.6\% lightbar) and acceptance (5.0\% eyes; 12.5\% lightbar). \edit{Although participants show noticeable individual variation in how strictly they use the rating scale, the aggregated results exhibit a consistent relative advantage of DesignerLLM over the base model across most metrics.} Commercial models achieve higher absolute ratings, but the improvements in our framework are statistically significant for most metrics, indicating that VLM-aligned training also yields measurable benefits for the quality of human-perceived eHMI.

\section{Conclusion}
We introduce \systemName, leveraging VLM perceptual feedback as an automated approach to progressively enhance the capabilities of LLM-based eHMI action designers. We perform extensive experiments and explore methods to effectively expand the action dataset while preserving alignment with VLM preference performance. Additionally, we conduct a user study that shows that insights from VLM raters can also improve human perception scores. Finally, our work serves as an important practice in improving the message delivery capabilities of LLM-actuated systems. In future work, we plan to incorporate real human feedback and richer multimodal teaching signals to better capture subtle and diverse human preferences that are difficult to approximate with VLM raters alone. 
%We also aim to improve the realism of the simulation environment by modeling more real-world visual factors and balancing rendering fidelity with efficiency for scalable training.

\section*{Limitations}

\paragraph{Collaborative Learning with Human Users.} The current setup demonstrates that, with the \systemName system, LLM-based eHMI action designers can achieve higher human perception scores, despite using VLM as an efficient substitute for real human annotators. However, as noted in previous work~\cite{xia2025automating}, the preference order between a human and a VLM is consistent only when the difference in the rating score between two actions is sufficiently large. Replicating the nuanced preferences of individuals from different cultural backgrounds or age groups remains a challenge. Therefore, incorporating a real human annotator into the refinement process would be valuable.

There are two aspects worth exploring. First, effective fine-tuning of the VLM rater models could further improve system performance. However, collecting enough data for a single round of fine-tuning is still costly, so the potential marginal gains from such fine-tuning merit careful consideration. Second, rating alone is often not an information-rich method for annotation. If we consider our system as a simulation of a real teacher, employing a teaching approach, using rich cues such as voice, gestures, posture, and other non-verbal signals, could serve as a more natural and effective way to teach LLM designers accurate scenario-action pairs. This approach could also help reduce the ambiguity associated with preference pairs, enabling more effective distillation of human preferences into the LLM training process.

\paragraph{Real-World Simulation and Learning.} In the current implementation (Section~\ref{method:data}), we utilize generated scenarios and rendered action videos as visual cues for VLM raters. This approach has proven to be both effective and efficient. However, it focuses mainly on simulating the main elements of typical traffic scenarios, such as automated vehicles, while overlooking other factors that could influence visual perception outputs for both VLM raters and human road users, such as lighting conditions, obstructions, and other environmental factors. Therefore, the challenge is to incorporate more visual elements that accurately simulate real-world scenarios in rendered videos while maintaining reasonable rendering times to enhance the robustness of LLM-based action designers in practical applications. One potential trade-off solution is to combine different levels of visual detail within the same training session. For example, including videos with rich visual elements for only a small portion of the training data could improve robustness while keeping overall training costs relatively low.

% \textbf{Efficiency-oriented iteration.} The current loop can be made faster by changing how samples are admitted to the expensive stages. One route is difficulty- or uncertainty-based scheduling: only actions whose VLM and HPM disagree, or actions for complex scenes, go through full evaluation and preference optimization; actions for simple, high-confidence scenes are evaluated once. A second route is staged evaluation: a light model (small VLM or the HPM) does fast filtering, and only top candidates are passed to the full evaluator. A third route is asset reuse: if the same scene and viewpoint are rendered once, later eHMI variants reuse the render instead of re-generating it. Together, these tactics reduce wall-clock time per round and make more co-learning iterations feasible under the same compute budget.

% \textbf{Improving evaluator reliability.} Even with a mixed evaluator, score drift can appear over long training runs. Periodically refreshing the HPM with a small new batch of human ratings, especially for failure patterns discovered by the VLM, can help maintain a stable evaluation space. Another option is to log evaluator disagreements and replay them as hard negatives during preference optimization. Both strategies keep the designer model aligned with human-facing judgments without requiring continuous human annotation.

\section*{Acknowledgments}
This work was supported by a Canon Research Fellowship and JST CRONOS, Grant Number JPMJCS24K8, Japan

\bibliography{reference}

\newpage
\appendix
% \section{Example Appendix}
\section{User Study Details}
\label{app:user_study}

In this section, we further evaluate the preference alignment of DesignerLLM models with human participants for two purposes: (1) to demonstrate that our co-learning framework can also enhance the perceptual experience of human users, and (2) to compare and analyze differences in preferences between VLMs and human raters, thereby exploring potential directions for future research.

\subsection{Study Design and Procedure}

\subsubsection{Experimental Setup.}
We selected two eHMI modalities (lightbar and eyes) as simple and common modalities for the user study. Eight traffic scenario–message pairs were adopted from prior work~\cite{xia2025automating} covering common interactions between AVs and other road users. Five LLMs were tested: two 7B-parameter models (DesignerLLM-base, trained after \textit{Format-aware Fine-tuning}, and DesignerLLM, after \textit{Iterative Preference-based Co-learning}) and three state-of-the-art models (OpenAI GPT-5, Claude Opus 4.1, and Grok 4). Each model generated eHMI actions that were rendered as 1080p (1980×1080) videos at 12 FPS to ensure smooth playback for participants in Blender. Unlike the videos provided to the VLM raters (Section~\ref{method:render}), these videos include rich contextual information. A total of 80 videos were created. Rendering all scenarios with detailed meshes and visual effects takes approximately 10 hours of computation.

\subsubsection{Interface Design.}
The user study was implemented and deployed using Gradio. Separate interfaces were created for the two eHMI modalities. Each interface consisted of five components:  
(1) a welcome page,  
(2) an introduction to general concepts of eHMI and the specific modality, including two demonstration videos,  
(3) a demographic information section,  
(4) main rating pages for eight pairs of scenarios and messages, and  
(5) an ending page.  
The interface included a session-based resume function, allowing participants to continue later using their assigned ID.

\subsubsection{Data Collection and Flow.}
\begin{figure*}[t]
    \centering
    \includegraphics[width=\linewidth]{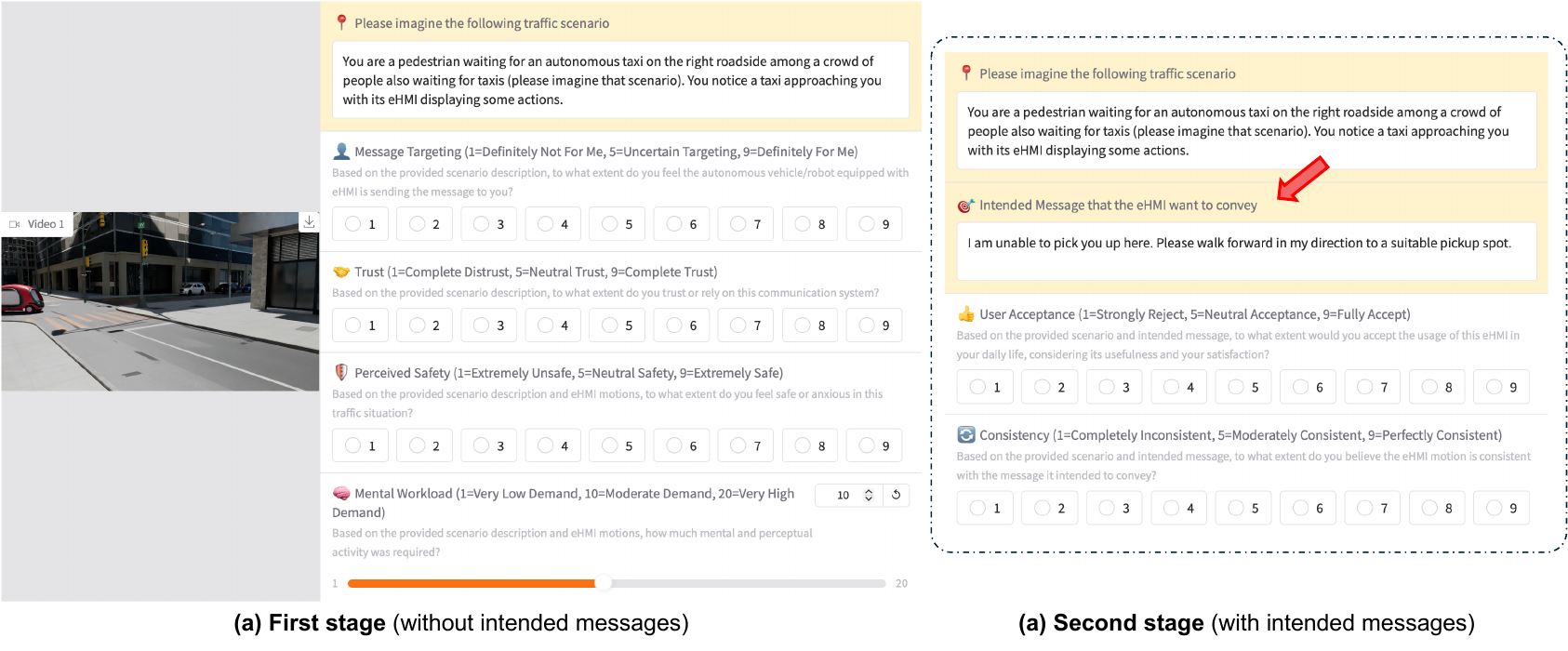}
    \caption{The layout for each video assessment is in two stages. In the first stage, participants rated their initial impressions of the eHMI using four metrics, without any reference to the intended messages. In the second stage, after the intended messages were revealed, participants re-evaluated their impressions using two different metrics.}
    \label{fig:user_study_page}
\end{figure*}

In the demographic section, participants anonymously reported their age, gender, and familiarity with eHMIs.  
For the main user rating pages, eight scenarios are evaluated across different pages, each presented in two stages: the first stage, where the intended messages are not provided, and the second stage, where the intended messages are included. This results in a total of 16 pages. On each page, five videos are shown in randomized order. As illustrated in \autoref{fig:user_study_page}, each video assessment follows a horizontal layout, with the video on the left and the questionnaires on the right. To ensure that no scores are missed, a verification mechanism is in place: participants must complete the current page before proceeding to the next, ensuring the completeness of the questionnaires.
Each session took approximately 5 minutes to complete.

\begin{figure*}[t]
    \centering
    \includegraphics[width=\linewidth]{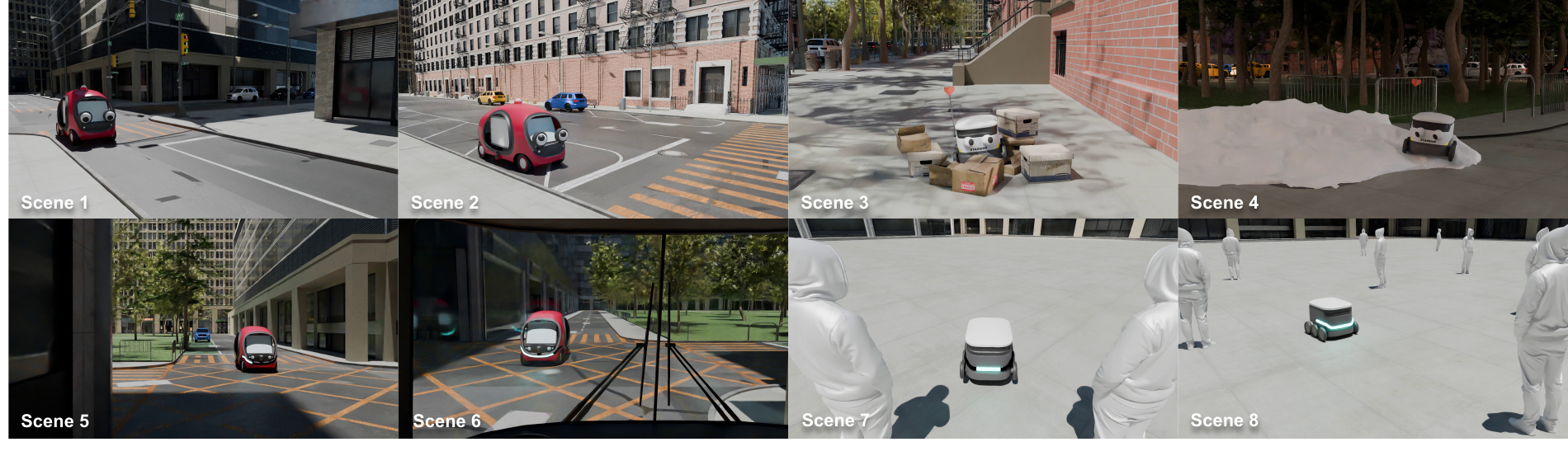}
    \caption{Demonstration of 3D scenarios used in the user study.}
    \label{fig:user_study_scene}
\end{figure*}

\subsection{Participants}
We recruited 18 participants aged 21 to 30 years (mean age: 26, SD = 2.7), consisting of nine males and nine females. They were evenly assigned to evaluate the lightbar and eye modalities. Among the participants, one was very familiar with eHMIs, having extensive knowledge or experience; two had some familiarity, having heard of eHMIs before; and the remaining fifteen were unfamiliar with eHMIs. All participants received a \$10 gift card as compensation.

\subsection{Measurements}
To assess the quality of eHMI actions generated by different LLMs, participants rated each video using standardized perceptual and cognitive metrics commonly used in eHMI studies~\cite{colley2020towards,colley2020design,colley2025improving}. Two measurement sets were used, depending on whether the intended message of the eHMI was provided.

\textbf{Without Intended Messages.}  
Participants rated their initial impressions of the eHMI using four metrics:
\begin{itemize}
    \item \textbf{Message Targeting (1–9):} The extent to which the eHMI message was directed towards the participant.
    \item \textbf{Trust (1–9):} The participant’s confidence in and willingness to rely on eHMI communication.
    \item \textbf{Perceived Safety (1–9):} The participant’s sense of safety or anxiety during the interaction.
    \item \textbf{Mental Workload (1–20):} The cognitive effort required to interpret eHMI actions.
\end{itemize}

\textbf{With Intended Messages.}  
After revealing the intended messages, participants reassessed using two metrics:
\begin{itemize}
    \item \textbf{User Acceptance (1–9):} The willingness to accept an eHMI in daily life.
    \item \textbf{Consistency (1–9):} The perceived alignment between the eHMI motion displayed and its intended message.
\end{itemize}

% \begin{figure*}[t]
%     \centering
%     \includegraphics[width=\linewidth]{figures/user_study_plot.pdf}
%     \caption{Human ratings of five models on six subjective metrics for (a) eyes and (b) lightbar. Bars show mean scores (1--9) with error bars; mental workload is rescaled from 1--20 to 1--9. Ranks are shown above bars (#1 best). DesignerLLM (7B) ranks 4th for both modalities, outperforming the base model and approaching commercial models. $\dagger$ denotes a simple eHMI description; $*$ denotes an enhanced prompt adapted from~\cite{xia2025automating}..}
%     \label{fig:user_study_result}
% \end{figure*}

\subsection{Result Analysis}
Prior to analysis, normality was assessed for these participant-level means using the Shapiro–Wilk test, and no serious deviations were found ($p > 0.05$) for either modalities. Because the study used a within-subjects design with five model conditions, each metric was analyzed using repeated-measures ANOVA.

\subsubsection{Reliability Analysis}

We assessed the reliability of subjective ratings before analyzing the effects of the model. Cronbach’s alpha was used to evaluate internal consistency within each metric across the five models. Internal consistency was excellent for both modalities (eye: $\alpha = 0.925$; lightbar: $\alpha = 0.953$), with all metrics exceeding $\alpha = 0.90$. Inter-rater reliability, estimated with the intraclass correlation coefficient (ICC), was low (eye: $ICC = 0.131$; lightbar: $ICC = 0.087$), indicating limited absolute agreement across participants. Because our analyses focus on relative comparisons within each subject in a single session, high internal consistency is the primary requirement; the low ICC limits generalization to settings that require stable scores between raters or sessions~\cite{koo2016guideline}.

\subsubsection{Eye Modality Performance}
For the eye modality condition with 9 participants, repeated measures ANOVA revealed significant main effects of the model type across all six metrics:

\textbf{Message Targeting.} ANOVA showed a significant main effect of model type ($F(4, 32) = 5.41$, $p = 0.002$, $\eta^2 = 0.403$). Grok~4 ($M = 5.47$, $SD = 0.89$) and OpenAI GPT-5 ($M = 5.47$, $SD = 0.88$) achieved the highest ratings, while the base model Qwen2.5-7B-Instruct showed the lowest performance ($M = 4.21$, $SD = 1.65$). Notably, \textbf{DesignerLLM} ($M = 4.86$, $SD = 1.21$) demonstrated a 15.4\% improvement over the base model, indicating that our co-learning framework successfully enhanced the message targeting metric despite its modest 7B parameter size.

\textbf{Trust.} A significant main effect was observed ($F(4, 32) = 4.66$, $p = 0.005$, $\eta^2 = 0.368$). Grok~4 ($M = 5.43$, $SD = 1.04$) achieved the highest trust ratings, followed by OpenAI GPT-5 ($M = 5.32$, $SD = 1.15$). \textbf{DesignerLLM} ($M = 4.65$, $SD = 1.32$) achieved a 7.9\% improvement over the base model ($M = 4.31$, $SD = 1.64$), demonstrating the framework's effectiveness in generating more trustworthy eHMI action design.

\textbf{Perceived Safety.} Model type significantly affected perceived safety ratings ($F(4, 32) = 4.82$, $p = 0.004$, $\eta^2 = 0.376$). Grok~4 ($M = 5.65$, $SD = 0.90$) and OpenAI GPT-5 ($M = 5.53$, $SD = 1.04$) achieved the highest ratings. \textbf{DesignerLLM} ($M = 5.00$, $SD = 0.89$) showed a 7.1\% improvement over the base model ($M = 4.67$, $SD = 1.06$), suggesting enhanced safety-oriented action generation through our framework.

\textbf{Mental Workload.} Analysis revealed significant differences in mental workload ($F(4, 32) = 4.23$, $p = 0.007$, $\eta^2 = 0.346$). Lower scores indicate better performance; Grok~4 ($M = 9.49$, $SD = 1.52$) and OpenAI GPT-5 ($M = 9.51$, $SD = 1.40$) imposed the lowest cognitive load, while the base model imposed the highest ($M = 10.97$, $SD = 1.38$). \textbf{DesignerLLM} ($M = 10.67$, $SD = 1.42$) demonstrated a 2.7\% reduction in mental workload compared to the base model, indicating more intuitive eHMI outputs.

\textbf{User Acceptance.} ANOVA indicated a significant main effect ($F(4, 32) = 4.23$, $p = 0.007$, $\eta^2 = 0.346$). Grok~4 ($M = 5.39$, $SD = 1.19$) and OpenAI GPT-5 ($M = 5.31$, $SD = 1.28$) achieved the highest acceptance ratings. \textbf{DesignerLLM} ($M = 4.38$, $SD = 1.48$) showed a 5.0\% improvement in acceptance over the base model ($M = 4.17$, $SD = 1.46$), validating the practical applicability of our framework's outputs.

\textbf{Consistency.} Significant differences emerged across models ($F(4, 32) = 3.73$, $p = 0.013$, $\eta^2 = 0.318$). Grok~4 ($M = 5.32$, $SD = 1.27$) and OpenAI GPT-5 ($M = 5.29$, $SD = 1.31$) achieved the highest consistency ratings. \textbf{DesignerLLM} ($M = 4.39$, $SD = 1.36$) exhibited 8.1\% better consistency than the base model ($M = 4.06$, $SD = 1.64$), demonstrating more reliable message-conduction capability.

Overall, \textbf{DesignerLLM} achieved an average rank of 4 across all metrics in the eye modality, substantially outperforming the base model (rank 5) with an average improvement of 7.7\% across all metrics while maintaining competitive performance relative to commercial models with significantly larger parameter counts.

\subsubsection{Lightbar Modality Performance}

For the lightbar modality condition with nine participants, repeated--measures ANOVA also revealed significant main effects for five of six metrics:

\textbf{Message Targeting.} ANOVA showed a significant main effect ($F(4, 32) = 6.34$, $p < 0.001$, $\eta^2 = 0.442$). Grok~4 ($M = 5.35$, $SD = 1.03$), OpenAI GPT-5 ($M = 5.31$, $SD = 1.08$), and Claude Opus 4.1 ($M = 5.32$, $SD = 0.95$) achieved the highest ratings. \textbf{DesignerLLM} ($M = 4.75$, $SD = 1.39$) demonstrated a 12.6\% improvement over the base model ($M = 4.22$, $SD = 1.33$).

\textbf{Trust.} A significant main effect was observed ($F(4, 32) = 4.40$, $p = 0.006$, $\eta^2 = 0.355$). OpenAI GPT-5 achieved the highest trust ($M = 5.53$, $SD = 1.09$), followed by Claude Opus 4.1 ($M = 5.50$, $SD = 0.97$) and Grok~4 ($M = 5.47$, $SD = 1.00$). \textbf{DesignerLLM} ($M = 5.04$, $SD = 1.17$) substantially outperformed the base model ($M = 4.67$, $SD = 1.20$) by 7.9\%, approaching the performance of commercial models.

\textbf{Perceived Safety.} Model type significantly affected safety perceptions ($F(4, 32) = 3.94$, $p = 0.010$, $\eta^2 = 0.330$). OpenAI GPT-5 ($M = 5.65$, $SD = 0.90$) and Grok~4 ($M = 5.60$, $SD = 1.02$) achieved the highest ratings. \textbf{DesignerLLM} ($M = 5.25$, $SD = 1.12$) showed a 7.1\% improvement over the base model ($M = 4.90$, $SD = 1.21$).

\textbf{Mental Workload.} No significant difference was observed across models ($F(4, 32) = 1.20$, $p = 0.331$, $\eta^2 = 0.130$). Nonetheless, OpenAI GPT-5 ($M = 9.47$, $SD = 1.90$) imposed the lowest cognitive load numerically. \textbf{DesignerLLM} ($M = 10.06$, $SD = 1.74$) showed a 3.8\% reduction in workload compared to the base model ($M = 10.46$, $SD = 1.83$).

\textbf{User Acceptance.} ANOVA indicated a highly significant effect ($F(4, 32) = 13.55$, $p < 0.001$, $\eta^2 = 0.629$), representing the largest effect size across all metrics and modalities. OpenAI GPT-5 ($M = 5.31$, $SD = 1.47$), Claude Opus 4.1 ($M = 5.29$, $SD = 1.29$), and Grok~4 ($M = 5.06$, $SD = 1.58$) achieved the highest acceptance ratings. \textbf{DesignerLLM} ($M = 4.33$, $SD = 1.66$) achieved a substantial 12.5\% improvement over the base model ($M = 3.85$, $SD = 1.75$), representing the largest single-metric gain.

\textbf{Consistency.} Significant model differences were found ($F(4, 32) = 7.54$, $p < 0.001$, $\eta^2 = 0.485$). OpenAI GPT-5 ($M = 5.40$, $SD = 1.39$) and Claude Opus 4.1 ($M = 5.28$, $SD = 1.31$) achieved the highest ratings. \textbf{DesignerLLM} ($M = 4.39$, $SD = 1.23$) demonstrated a 3.5\% improvement in consistency over the base model ($M = 4.24$, $SD = 1.15$).

In the lightbar modality, \textbf{DesignerLLM} maintained an average rank of 4.0, consistently outperforming the base model (rank 5.0) across all metrics with an average improvement of 7.9\%, demonstrating cross-modality robustness of our co-learning framework.

% \subsubsection{Cross-Modality Consistency}

% \textbf{DesignerLLM} exhibited perfect rank consistency across both modalities (rank 4.0 in both eye and lightbar conditions), indicating robust and generalizable performance improvements from our framework. This consistency surpassed all commercial models, which showed ranking variations between modalities (Grok~4: rank 1.0 in eye vs.\ rank 2.3 in lightbar; OpenAI GPT-5: rank 2.0 in eye vs.\ rank 1.3 in lightbar).

\subsubsection{Summary}

The human evaluation results demonstrate that our co-learning framework successfully enhances a 7B parameter model (\textbf{DesignerLLM}) to generate significantly more acceptable eHMI actions compared to its base model counterpart (Qwen2.5-7B-Instruct). Across both modalities, \textbf{DesignerLLM} achieved an average improvement of 7.8\% across all metrics, with the largest gains observed in message targeting (15.4\% for the eye and 12.6\% for the lightbar) and user acceptance (5.0\% for the eye and 12.5\% for the lightbar). Although commercial models with substantially larger parameter counts, such as GPT-5, Grok~4, and Claude Opus 4.1, achieved higher absolute ratings, \textbf{DesignerLLM} consistently ranked fourth across both modalities and all six metrics, validating the effectiveness of our framework. Notably, the large effect sizes observed across most metrics ($\eta^2$ ranging from 0.318 to 0.629) indicate that model choice has a substantial impact on perceived eHMI quality. These results confirm that our framework not only improves automated VLM-based evaluation scores but also translates into meaningful enhancements in real human user experience, achieving competitive performance with a fraction of the computational resources required by commercial alternatives.

% Reliability check – mention whether Cronbach’s alpha or inter-rater reliability was calculated.

% Aggregation – clarify how the scores are averaged (per participant, per model, or per scenario).

% Justification – cite prior studies that use similar perceptual or trust metrics (e.g., SAE J3016 or eHMI trust studies).

% Optional figure – include an example questionnaire layout screenshot for clarity.

% Statistical handling – specify later how differences between models were tested (e.g., ANOVA or mixed-effects models).

\edit{
\section{Evaluator Prompts and Templates}
\label{app:prompts}

This appendix provides the full prompts used in the multi-metric evaluation system (Section~3.3). All evaluations use GPT-5-mini~\cite{openai2025gpt41}. The video input consists of up to 50 frames, each resized to $512\!\times\!512$ pixels.

\subsection{Phase~1: Pre-Evaluation Prompt (Without Intended Message)}
\label{app:phase1}

\subsubsection{System prompt}

\begin{promptbox}
\small
You are a communication receiver in a traffic condition.

The detailed situation you encounter is as follows:
\texttt{\{receiver\_scenario\}}

You are conducting an experiment to evaluate the message conveyance capability of the eHMI action (movement). You can clearly notice there is an external Human-Machine Interface (eHMI) installed on this autonomous vehicle.

The eHMI type you see installed on the car is:
\texttt{\{ehmi\_short\_description\}}

\textbf{What You Will Receive:}
\begin{itemize}[nosep]
    \item \textbf{Video frames} showing eHMI behavior and animations. The video is rendered in Blender, but please treat it as a real-life scenario. The video will provide correct visual effects, lighting, and viewing angle from your perspective. The background is removed for fast rendering, so please ignore background effects and context when making your decision, focusing only on the action (movement) of the eHMI.
    \item \textbf{Specific evaluation questionnaires} to guide your assessment.
\end{itemize}

\textbf{What You Will Not Receive:}
\begin{itemize}[nosep]
    \item \textbf{The intended message} that the vehicle equipped with the specific eHMI intends to convey to you. You are completing this questionnaire without knowing the intended message to ensure an unbiased evaluation based purely on your natural interpretation of the eHMI behavior.
\end{itemize}

\textbf{Your Role and Approach.} You should:
\begin{itemize}[nosep]
    \item \textbf{Act as a real human receiver} in a traffic condition, considering factors that might affect human raters, including but not limited to: visual attractiveness, clarity, timing, message effectiveness, friendliness, and safety perception.
    \item \textbf{Fully immerse yourself} in the provided scenario, thinking about what a human communication receiver would think and need in this situation.
    \item \textbf{Provide your immediate reaction} as you would naturally respond in real traffic conditions.
    \item \textbf{Give detailed reasoning} for your ratings, explaining what specific elements influenced your assessment.
    \item \textbf{Consider real-world implications} of how this eHMI communication might affect traffic safety and human behavior.
\end{itemize}
\end{promptbox}

\subsubsection{User prompt}

\begin{promptbox}
\small
Based on the scenario and provided eHMI action (movement) video, please analyze what message you believe the eHMI is trying to convey, assess whether this communication is directed at you, and evaluate your trust in this autonomous vehicle system.

\textbf{Intention Recognition.}
Please carefully observe the eHMI behavior and infer what message the autonomous vehicle is attempting to communicate to you as a traffic participant. Consider the following aspects: visual patterns, movements, and animations; timing and duration of the eHMI behavior; color changes, brightness, or other visual cues; the context of the traffic scenario; what action or understanding the vehicle might be seeking from you; vehicle's relative position and movement direction in relation to your location; and whether the eHMI signals appear to be directed specifically at you or other traffic participants.

Rate your certainty in your message interpretation on a 9-point scale:
\textbf{1}~=~completely uncertain $\to$ \textbf{5}~=~neutral certainty $\to$ \textbf{9}~=~completely certain.

\textbf{Message Targeting Assessment.}
Please assess whether you believe the eHMI communication is specifically directed at you as a traffic participant.

Rate your confidence on a 9-point scale:
\textbf{1}~=~definitely not for me $\to$ \textbf{5}~=~uncertain targeting $\to$ \textbf{9}~=~definitely for me.

\textbf{Trust Evaluation.}
Please evaluate the degree to which you believe the autonomous vehicle and its eHMI are reliable and will act safely as indicated.

Rate your trust on a 9-point scale:
\textbf{1}~=~complete distrust $\to$ \textbf{5}~=~neutral trust $\to$ \textbf{9}~=~complete trust.
\end{promptbox}

%% ---
\subsection{Phase~2: Post-Evaluation Prompt (With Intended Message)}
\label{app:phase2}

\subsubsection{System prompt.}
The setup is identical to Phase~1 (Appendix~\ref{app:phase1}), except the clause \textit{``What You Will Not Receive''} is replaced with \textit{``What You Will Receive''}.

\subsubsection{User prompt.}

\begin{promptbox}
\small
The intended message is as follows:
\texttt{\{intended\_message\}}

You are provided with this message after you experience this scenario (watch the video).

Based on the scenario, provided eHMI action (movement) video, and the intended message the eHMI action wants to convey, please answer the following questions:

\textbf{User Acceptance Evaluation.}
Please evaluate the degree to which you would accept and approve of this eHMI-equipped autonomous vehicle based on the behavior shown in the video frames.

Rate the overall user acceptance on a 9-point scale:
\textbf{1}~=~strongly reject $\to$ \textbf{5}~=~mixed acceptance $\to$ \textbf{9}~=~fully accept.

\textbf{Consistency Evaluation.}
Please evaluate the consistency between the eHMI actions shown in the video frames and the intended message described above.

Rate the overall consistency on a 9-point scale:
\textbf{1}~=~completely inconsistent $\to$ \textbf{5}~=~mixed consistency $\to$ \textbf{9}~=~perfectly consistent.
\end{promptbox}

%% ---
\subsection{Similarity Judge Prompt}
\label{app:similarity}

After Phase~1, the VLM-interpreted message is compared with the ground-truth intended message using a small LLM judge (GPT-5-mini).

\begin{promptbox}
\small
You are evaluating the similarity between an intended message and an interpreted message from an eHMI (external Human-Machine Interface) interaction.

\textbf{Task:} Compare the intended message (ground truth) with the interpreted message (what the VLM guessed) and provide a similarity score.

\textbf{Scoring Criteria.}
Focus on the main meaning and urgency level:
\begin{enumerate}[nosep, leftmargin=*]
    \item \textbf{Main Action Similarity:} Do both messages convey the same core action/intention? (e.g., yielding, stopping, proceeding, waiting, warning)
    \item \textbf{Urgency Level:} Do both messages show similar urgency or immediacy?
    \item \textbf{Overall Meaning Alignment:} Do the messages fundamentally communicate the same thing to a traffic participant?
\end{enumerate}

\textbf{Input:}

Intended Message: \texttt{\{intended\_message\}}

Interpreted Message: \texttt{\{interpreted\_message\}}
\end{promptbox}

%% ============================================================
\section{Training Hyperparameters}
\label{app:hyperparams}

All fine-tuning experiments use LLaMA-Factory~\citep{zheng2024llamafactory} with LoRA~\citep{hu2022lora} in bfloat16 precision. The data are formatted in ShareGPT style~\citep{vicuna2023} and split into training and test sets at an 80/20 ratio. Table~\ref{tab:hyperparams} lists the full configuration for both stages.

\begin{table}[h]
\centering
\small
\caption{Training hyperparameters for Format-aware Fine-tuning (SFT) and Iterative Preference-based Learning (DPO).}
\label{tab:hyperparams}
\begin{tabular}{lcc}
\toprule
\textbf{Parameter} & \textbf{SFT} & \textbf{DPO} \\
\midrule
Base models             & \multicolumn{2}{c}{Qwen2.5-7B / 1.5B-Instruct} \\
Fine-tuning method      & LoRA  & LoRA  \\
LoRA rank               & 8     & 8     \\
Learning rate           & $5.0\!\times\!10^{-5}$ & $5.0\!\times\!10^{-6}$ \\
LR scheduler            & cosine & cosine \\
Warmup ratio            & 0.1   & 0.1   \\
Epochs                  & 3     & 1     \\
Effective batch size    & 16    & 16     \\
Cutoff length           & 2048  & 2048  \\
Precision               & bf16  & bf16  \\
\midrule
\multicolumn{3}{l}{\textit{DPO-specific}} \\
\quad DPO $\beta$            & --- & 0.2 \\
\quad DPO loss             & --- & sigmoid \\
\quad SFT mixing weight       & --- & 0.01 \\
\bottomrule
\end{tabular}
\end{table}

\paragraph{Preference pair construction.}
For each scenario, pairs $(y^{+}, y^{-})$ are formed using the kernel score $K$. We first select the max--min pair when $\Delta K_{i} \geq \delta_{\min}$, where the default 
$\delta_{\min} = 4.0$. 
We then rank the remaining high-gap pairs within the same scenario by score difference and retain the top 30\%.

\paragraph{Quantization for local inference.}
During DPO candidate generation, models are loaded with 4-bit NormalFloat (NF4) quantization.

}
\begin{figure*}[h]
    \centering
    \includegraphics[width=\linewidth]{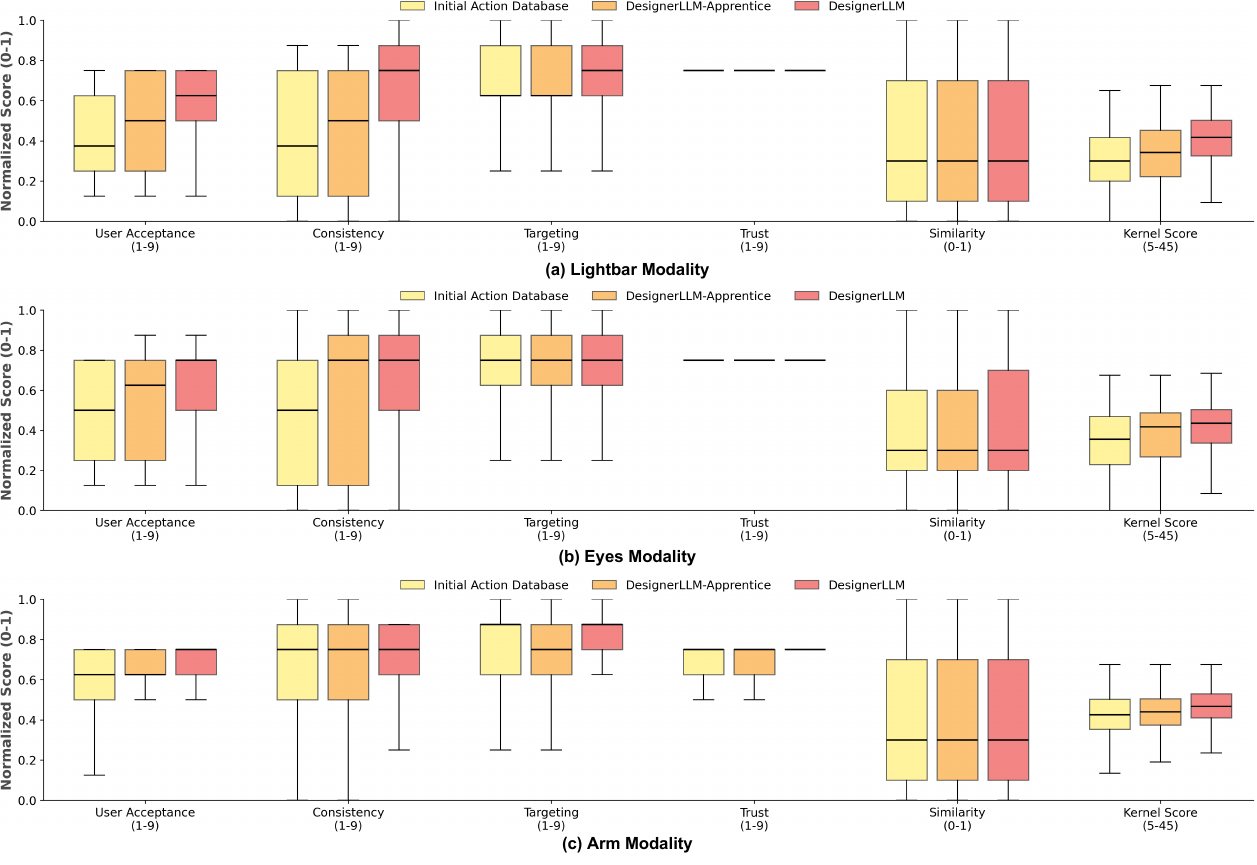}
    \caption{Boxplots of User Acceptance, Consistency, and Trust across training milestones. User Acceptance and Consistency increase over time and show reduced variability, while Trust remains largely unchanged.}
    \label{fig:vlm_metric_plot}
\end{figure*}

% \section{Example Appendix}
% \label{sec:appendix}
\end{document}